\documentclass[12pt,preprint]{aastex}

\usepackage{epstopdf}
\usepackage{graphicx}

\def\opeqn{\begin{equation}}
\def\cleqn{\end{equation}}
\def\lsim{\hbox{\rlap{\raise 0.425ex\hbox{$<$}}\lower 0.65ex\hbox{$\sim$}}}
\def\gsim{\hbox{\rlap{\raise 0.425ex\hbox{$>$}}\lower 0.65ex\hbox{$\sim$}}}

\begin{document}

\title{A Thought of Wrapping Space Shuttle External Tank
         with Ceramic Fiber Fishnet Stockings} 
\author{Sun Hong Rhie}


\begin{abstract}

The new camera system of the shuttle Discovery on STS-114 that blasted off 
at 10:39am, Tuesday, July 26, 2005, after 906 days of grounding since the 
Columbia accident, has produced high resolution data of foam sheddings.
The $0.9$ lbs piece from the Protuberance Air Load (PAL) ramp 
on the LH2 tank is believed to be comparable in its potential adversities 
to the $\sim 1.67$ lbs BX-250 foam from the $-Y$ bipod ramp that demised 
shuttle Columbia in 2003. The two known incidences indicate that 
protuberant foams, possibly in conjunction with the liquid hydrogen 
temperature, offer lame targets of the aerodynamic forces. 
Seven other relatively large divots in the STS-114 external tank foam 
insulation have been reported, and foam shedding remains to be a challenge
to be resolved before the next space shuttle launch. 
The relatively large divots from the newly streamlined foam around the
-Y bipod area suggests a potential necessity for a new line of resolution. 

We suggest an option to wrap the insulated external fuel tank with a grid of 
high temperature resistant ceramic fibers 
({\it ceramic fiber fishnet stockings}).  
Assuming fiducial acreage of $20000 ft^2 $, 
one inch square cell single fiber grid will weigh only $60g$
with fiber cost \$66. Even with 1500-fiber-equivalent strength, 
one inch square cell grid will add only 
$200\,lbs$ and ``miniscule" \$100,000.

\end{abstract}

\maketitle


\section {Foam Losses on STS-114}
\label{secFoamLoss}

While HST customers were crossing their fingers, the STS-114 
camera system exhibited a spectacular scene of a top-hat shape $0.9 lbs$  
insulation foam flying off the external tank.
The public release of the foam loss locations
 by NASA shown in Fig.\,1 read as follows.
\begin{enumerate}
 \item
  LH2 PAL ramp: $36.3^{''} \times 11^{''}$ section near stattion Xt-1281
released at 127.1 sec MET;
 \item
  $8.4^{''} \times 7.3^{''}$ divot on the forward outboard portion of 
the -Y bipod closeout at 148.1 sec MET;
 \item
  Divot ($\sim 5.6^{''}L \times 3.5^{''}W \times 2.9^{''}D$) 
in LH2 tank ice/frost ramp at Xt-1262 at 154.8 sec MET 
(exposed Conathane adhesive at center of divot);
 \item
 Divot ($\sim 7.3^{''}L \times 1.9^{''}W \times 2.5"D$) in LH2 tank 
ice/frost ramp at Xt-1525 (exposed Conathane adhesive);
 \item
 Shallow divot ($\sim 4.0^{''}L \times 2.6^{''}W \times 0.3^{''}D$) 
in LH2 tank ice/frost ramp at Xt-1841;
 \item
 Divot ($\sim 4^{''}$ dia) $\sim 3 $ft below the -Y bipod fitting
(Xt-1163) at 135.8 sec MET;
 \item
 Shallow divot ($\sim 8^{''}-10^{''}$ dia) in LH2 tank acreage between 
LO2 feedline and pressline (Xt-1839);
 \item
 2 divots ($\sim 3^{''}$ dia) on upper intertank/LH2 tank flange closeout 
on -Y axis (release time unknown).
\end{enumerate}

\section {``Rationale" for an Alternative Resolution}

The press release by NASA immediately following the dislodge of the large 
$0.9$ lbs piece of insulation foam from PAL ramp indicates NASA's continuing 
efforts to prevent defects in foam bubbles and
to tame supersonic vortices and turbulences as it reads in the last 
paragraph: ``... An enhanced spray process is in
work for future tanks, as well as continued work in
developing redesign options including elimination of the ramps; reducing
the ramps' sizes by two thirds; or building a trailing edge `fence' on the
back side of the cable tray, which would act like a nozzle throat and prevent
unsteady flow in that area."

The divots in items 2 and 8 in section\,\ref{secFoamLoss} are from  
the left side (-Y) of the bipod area. This is exactly the same
area where the $\sim 1.67$ lbs piece of foam broke off from the ramp and
doomed shuttle Columbia in 2003. The ramps have been removed, 
and the surface of the foam in the bipod area has been streamlined since.
But, the air managed to dump enough 
energy to break off large chunks from the streamlined foam. 
The fourth volume of the Columbia Accident Investigation 
Board (CAIB) Report\footnote{
``ET Cryoinsulation" in CAIB Report Volume IV: Appendix F.4, p.16; 
See also Fig.\,3.2.5 in Volume I; http://www.caib.us.} exhibits  
a result of hydrodynamic simulations that shows streamline vortices around 
the -Y bipod area. We have no clue, especially not being an expert either
in supersonic hydrodynamics or in external tank configurations, 
whether the vortices around the obstacles of the
struts that hold the nose of the shuttle can be tamed without undesirable,
or unpredictable and untested, side effects.  However, NASA's 
investments of 2.5 years and $\sim 1$ billion dollars since Coumbia accident 
point to a possibility that the same line of efforts 
cited in the first paragraph of this section
might have already served its capacity. 
   
An alternative is suggested to strengthen the insulated external fuel tank 
with fish-net stockings made of high temperature resistant 
ceramic fibers that are immediately available. Oxide fiber 
3M Nextel 720 manufactured by 3M Technologies is considered.

\section {No Foam Loss as Specified in the Book}

\begin{paragraph}
\noindent
{\bf 3.2.1.2.17 External Tank Debris Limits:} {\it No debris shall 
emanate from the critical zone of the External Tank on the launch pad
or during ascent except for such material which may result from normal
thermal protection system recession due to ascent heating\footnote{
\rm ``External Tank End Item (CEI) Specification -- Part 1," CPT01M09A,
contract NA-58-30300, April 9, 1980, WBS 1.6.2.2.}.}  
\end{paragraph}

It is the design requirements 
in ``Flight and Ground System Specification-Book 1, Requirements" 
as quoted in CAIB report volume 1 chapter 6, p.122.
The report continues, ``The assumption that only tiny pieces of debris
would strike the Orbiter was also built into original design requirements,
which specified that the Thermal Protection System (the tiles and RCC
panels) would be built to withstand impacts with a kinetic energy less
than $0.006$ foot-pounds." 

\begin{figure}
 \plotone{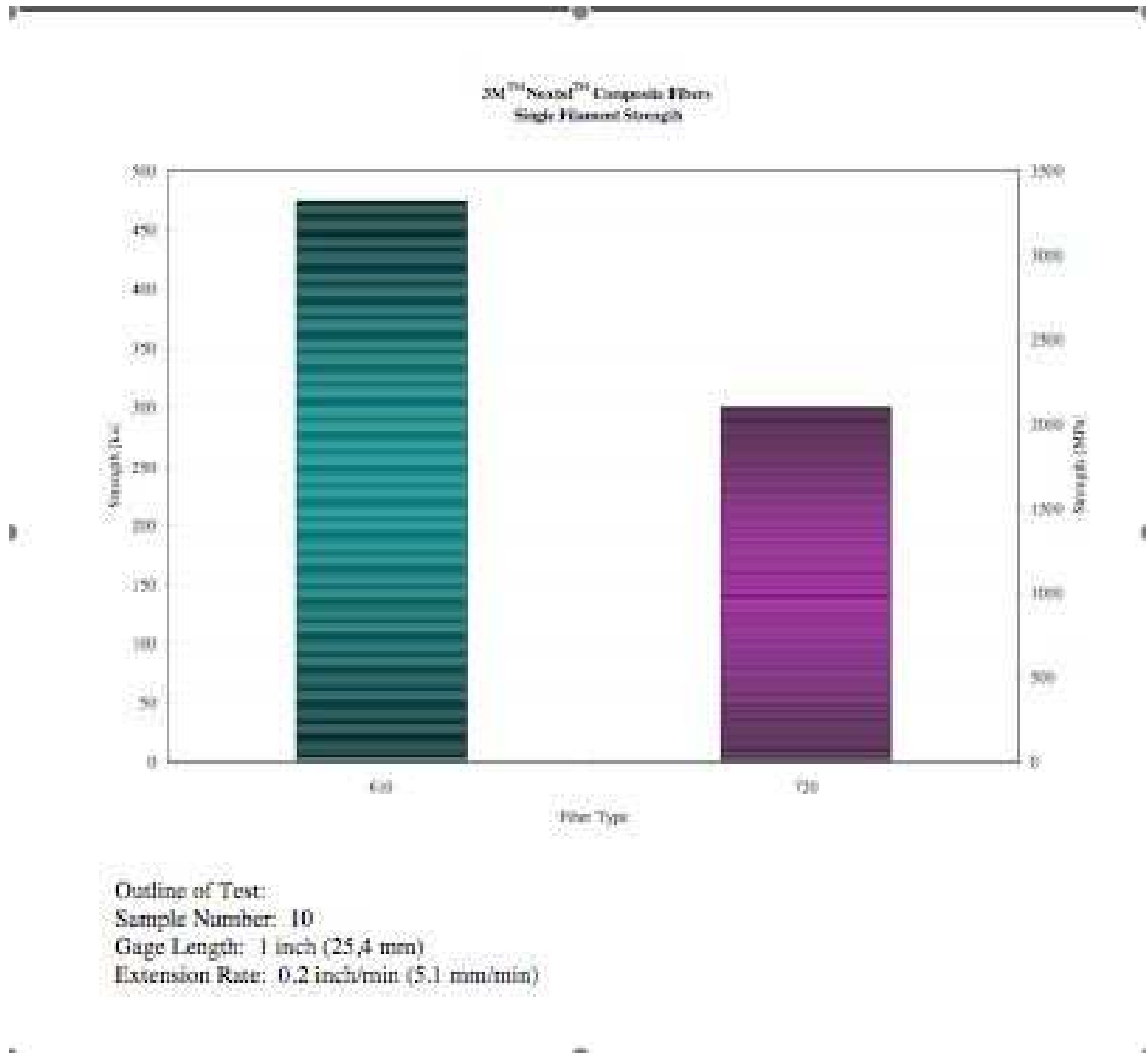}
 \caption{ 
Room temperature single fiber tensile strengths of fiber 610 and fiber 720. 
(The figure is from 3MTTN.)}
\label{fig_singlefiber}
\end{figure}

The corresponding maximum mass and size of the 
allowed foam bebris is $0.652 g$ and $\sim 1 in^3$ where the latter is 
commonly referred to as ``the size of a marshmallow" in news media. The large
piece foam debris from the PAL ramp on STS-114 is $\sim 620$ times the maximum 
allowed. 
Nextel 720 $\alpha$-$Al_2 O_3$ fiber properties shown in figures 
\ref{fig_720roving}, \ref{fig_singlefiber}, and \ref{fig_heating90sec}  
(electronically cut out of 3MTTN) indicate that the ceramic fiber technology 
is sufficiently mature to meet the design specification of ``no foam loss".

\begin{enumerate}
 \item
In civilized units (cgs or mks), the impact energy limit 
$0.006$ foot-pounds is about $8.13 \times 10^{4} ergs$.
Some of the convesion formulae for pressure (or strain) are listed.
\opeqn
 1 Pa(scal) \equiv 1 Newton/m^2 = 1 kg\, m^{-1} sec^{-2}
       = 1.4503774 \times 10^{-4}\,psi
\cleqn
\opeqn
 1 atm = 101325 Pa = 1.01325 bar = 14.6959\,psi
\cleqn

 \item
In order to convert the impact energy limit into the maximum allowed mass 
of a foam debris, we need to know the areodynamic velocity due to the ascending
motion of the spacecraft and wind, air density, streamline, etc,  
as a function of time and the position in the external tank.  
A recent review, found in the web site of Lockheed Martin, 
``STS-114 Flight Readiness Review: External Tank (ET-121),"
Appendix D, P.5, June 29-30, 2005,  
indicates that the mass of a foam debris should not exceed 
$\sim 0.023 oz = 0.652 g$. 
The density of the foam is $\sim 2.4 lbs/ft^3$, henced the volume of the 
infimum mass foam debris is $\sim 16.96 cm^3 \approx 1 in^3$.
We take $0.652g$ and $1in^3$ as the currently accepted mass and volume limits. 

 \item
Table 7.2.1.8.1.2.6-1 of CAIB Report titled ``Summary of Acreage Stress
Analysis Parameters" lists flight requirement (foam) cell pressures in various 
parts of the foam insulation on the external tank: 
Oxygen tank: $23.6$, $19.2$, $15.9\,psi$ depending on the parts;
Intertank: $19.7\,psi$;
Hydrogen tank: $16.3\,psi$, and bond tension test requirement: $30.0\,psi$.
Hence we take $30\,psi$ as the tensile strength requirement of the 
ceramic fiber grid structure. 

\end{enumerate}

\subsection{High Temperature Resistant Ceramic Fiber 3M Nextel 720}
 
High tensile strength fibers are based in carbon, SiC, and 
$Al_2 O_3$\footnote{ ``High Temperature Structural Fibers -- Status and Needs", 
J.A.DiCarlo, NASA TM-105174, 1991.}.
 Space shuttle RCC panels -- now a household name since the Columbia accident 
-- are made of carbon fiber composite matrix with SiC protection coating and 
glassy sealant\footnote{ ``Carbon Fiber Composites", D.D.L.Chung, 
Butterworth-Heinmann, 1994.}${}^,$\footnote{
``Carbon-carbon composites: engineering materials for hypersonic flight", 
NASA TM-103472, 1989.}.
 During the reentry of the shuttle, the sealant melts forming an air tight 
thin film. A report\footnote{
`` Modeling the Thermostructural Capability of Continuous Fiber-Reinforced 
Ceramic Composites", J.A.DiCarlo and H.M.Yun, Journal of Engineering for 
Gas Turbines and Power, 2002, p.467: DY hetherfrom.} indicates
that oxide fiber tows (or roving: loose bundle of untwisted fibers) perform 
better than SiC-based tows in the condition under consideration where 
it is highly oxidating and the usage time is short ($q \,< \, 30000$). 
The interactions within the bundle make the tow weaker or stronger in terms 
of creep or rupture in comparison to the average strength of 
the single fiber. At $1200^\circ C$ , single fiber SiC outperforms SiC tow; 
At $1400^\circ C$, SiC tow outperforms single fiber SiC. 
We consider an oxide fiber 3M Nextel 720. 

3M Nextel 720 fiber\footnote{
``3M Nextel Ceramic Textiles Tecnical Notebook" (http://www.3m.com): 
3MTTN from hereon.}
is a continuous fiber with the composition of 85\% $Al_2 O_3$ and 
15\% $SiO_2$, melting temperature $1800^\circ C$ , density $3.40g/cm^3$, 
the filament tensile strength $2100MPa (300ksi)$ at $25.4mm$ guage, 
the diamter $\sim 12 \mu m$, and the crystal phase $\alpha$-$Al_2 O_3$+mullite. 
Nextel oxide fibers are various mixtures of $Al_2 O_3$, $SiO_2$, and $B_2 O_3$. 
Fiber 610 is made of pure $Al_2 O_3$ ($> 99\%$) and its crystal phase is
$\alpha$-$Al_2 O_3$. 

\begin{figure}
 \plotone{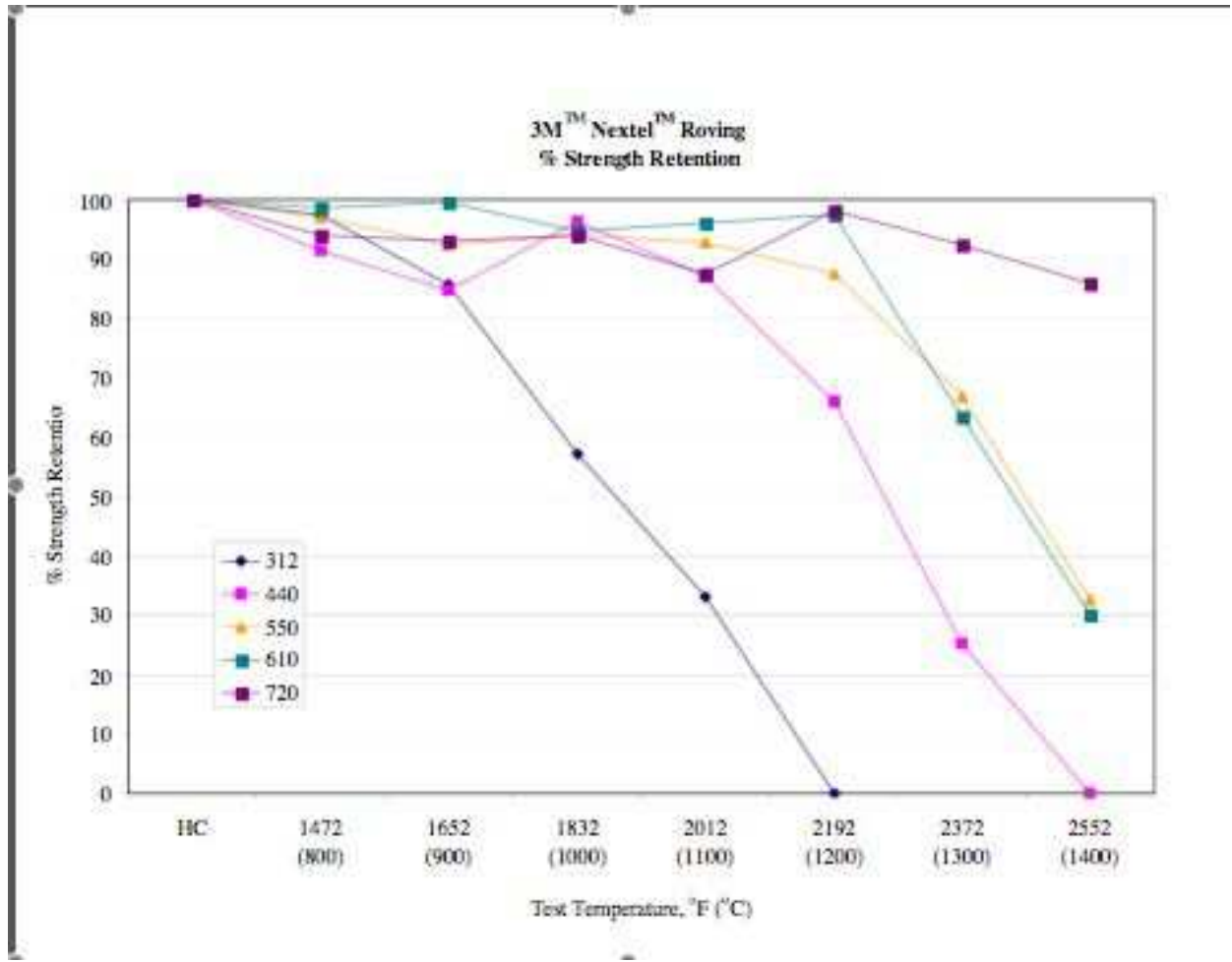}
 \caption{ 
3M Nextel 720 tow shows relativley high stability in high temperatures. Fiber 
610 shows sudden drop in strength at $1200^\circ C$ . (The figure is from 3MTTN.)}
\label{fig_720roving}
\end{figure}

\begin{figure}
 \plotone{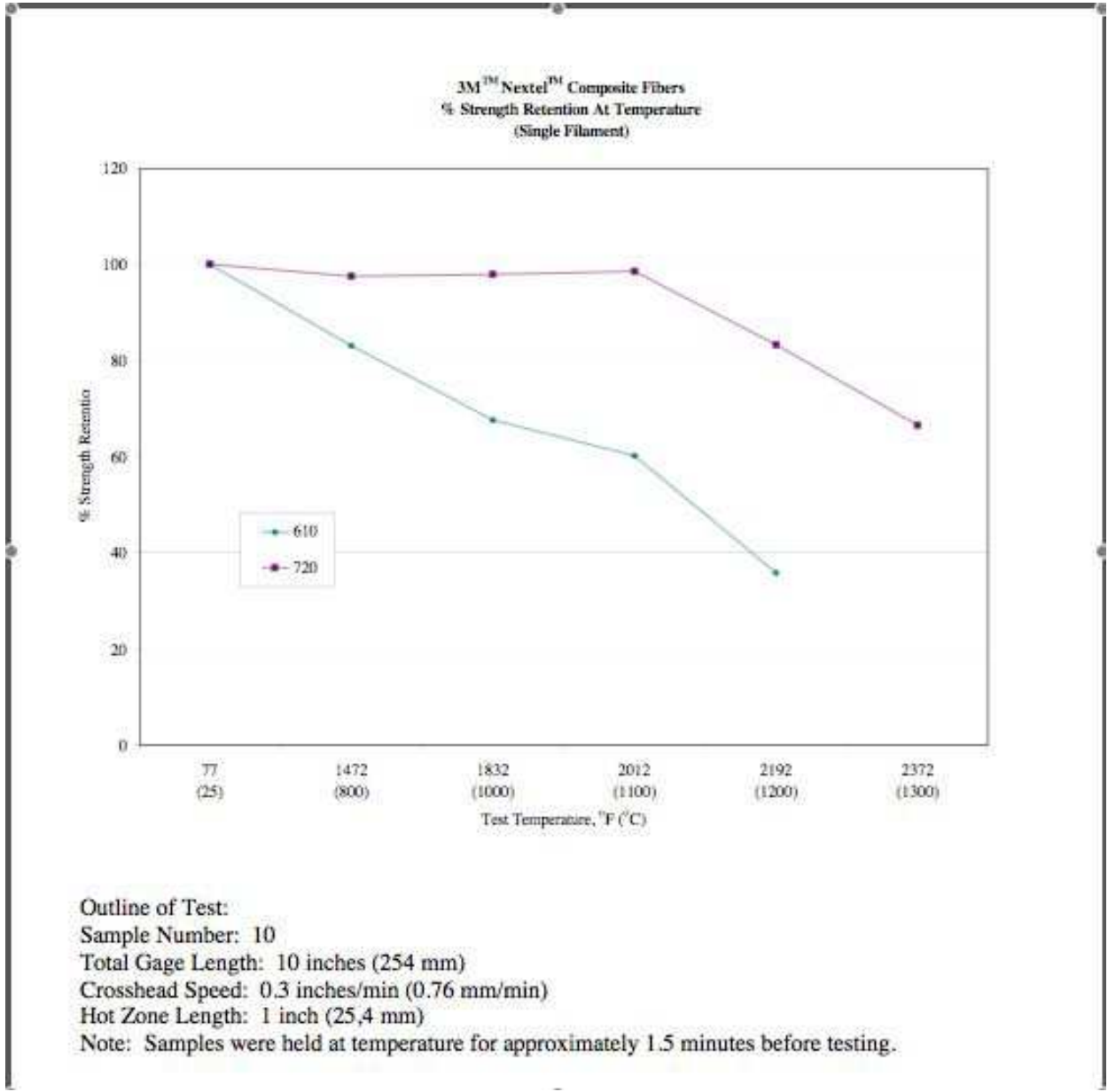}
 \caption{ 
Fiber 720 shows a relatively stable strength at high temperature 
heating for $\sim 90 sec$. The maximum aerodynamic heating of the external
tank occurs at $\sim 80 sec$ after the launch. (The figure is from 3MTTN.)}
\label{fig_heating90sec}
\end{figure}

Figure \ref{fig_720roving} shows the stable 
strength of fiber 720 tow in high temperatures. 
(3M Nextel oxide fiber tows come in fiber counts of 400 or 750.) 
Figures \ref{fig_singlefiber} and \ref{fig_heating90sec} show the single fiber 
tensile strengths of the fibers 610 and 720 in room temperature and high 
temperatures where the high temperature heatings lasted about 90 seconds. 
(Note that the maximum aerodynamic friction and heating occurs around $80$
seconds after the launch.) The fiber 610 is stronger at room temperature, 
but fiber 720 is preferred because of the stability of the strength in high 
temperatures.  Comparision of figures \ref{fig_720roving} and 
\ref{fig_heating90sec} shows that the fiber 720 tow performs better than 
single fiber at higher temperatures ($ > 1150^\circ C$) consistently with 
the report by DY. 

The concern of foam shedding is on the area where the aerodynamic stress is 
high, and the design requirement\footnote{
``Return to Flight Focus Area: External Tank Thermal Protection System", 
www.nasa.gov.} of the temperature tolerance due to 
aerodynamic heating is $1200^\circ C$ .  Within the temperature range, 
fiber 720 shows excellent strength behavior and also very 
slight difference in the strengths between a tow and an average single fiber. 
The temperature tolerance requirements for the areas with heating from solid 
rocket booster and from main engine are very high with $1650^\circ C$ 
and $3300^\circ C$. Foam shedding or ablation from those areas 
(aft dome and lower part of the acreage) is not a concern in regard to the 
potential damage to the thermal protection systems (TPS), and reinforcement 
fibers should not be used to avoid generating unnecessary potential hazard 
of loose fiber fragments or melts.

\subsection { Added Mass Estimation }
 
\begin{enumerate}
 \item
The total acreage of the external tank foam insulation is $\sim 24000 ft^2$. 
(The total mass of the foam is $\sim 4800lbs$.) 
The area that needs fiber reinforcement is likely to be less than $20000 ft^2$, 
and $20000 ft^2$ will be considered the fiducial acreage. Most of the foam 
loss seems to come from intertank area where liquid oxygen tank and liquid 
hydrogen tank are joined (CAIB report). 

\item
Consider a single fiber one inch square cell grid. The number of cells in the grid 
is $2,880,000$, the total length of the fiber in the 
grid is $5,760,000\,in$, and the total mass is $60\,g$. 
(According to 3MTTN, 400 fiber count 
tow has 1500 denier (denier = number of grams in 9000\,m or 10000\,yd of 
a product).)
Assuming absolute no failure of the single fiber grid, added weight of 
$60\,g$ of fibers will keep the foams from being released. It is most likely 
not practicable, but the negligibly small mass of the single fiber grid is 
a good news. The cost of the single fiber grid will be \$66. 
It is only getting better. 

\item
The tensile strength measurement is done with a short piece of 
fiber of length $25.4\,mm$, and I did not find failure probability 
of long fibers in 3MTTN. The total strength equivalent of 400 or 750 
fibers of the commercial products may be more than fail-safe for the 
lengths of the girth and height of the appication area of the external tank. 
One can consider woven yarns or very narrow fabric to replace the single 
fiber strand for the grid.  The ceramic fiber net may be embedded
near the surface of the foam or wrap the surface.
A 1500 fiber-equivalent strength grid 
will add mass $\sim 200\,lbs$ with the price tag of \$100,000. 

\end{enumerate}

The analysis is preliminary and the mass estimation is made 
over a broad range 
because of the lacking information of the properties of long fibers,
exact behavior of the insulation foam under stresses, and so the
engineering formulation of the grid. For example, the foam will have 
surface waves generated by the aerodynamic flows, and the fiber (yarn) would 
need to have certain elasticity whether the fiber grid is embedded or
on the surface.\footnote{J.D.Walker indicates in
``Impact Modeling", CAIB Report, Volume II, D12 that the foam behaves 
elastically with strain rates higher
than $7\times 10^{-3}/sec$ where the estimated Young modulus of the foam 
is $8.0 Mpa$.}
 The proper elasticity may be engineered through the 
bundle structure such as braiding or weaving. 
Inelastic fiber can act as a ``diamond string saw" to an adverse effect. 
Even if it shaves off only benignly small pieces, it should be best to 
avoid generating additional foam flakes.     

In summary, 
a ceramic-fiber-grid reinforcement can eliminate foam-shedding problem.
The projected mass and price of the grid are small. 
There are some details to be specified to engineer a  
practical ceramic-fiber grid that will secure insulation 
foam of the external tank -- economically.  
``No foam loss" makes the flight safer and reduces turn-around repair costs. 
We call for cooperations from the interested parties.

\vskip 1.0truecm
\centerline{\bf Acknowledgments}

We appreciate discussions with J.A.DiCarlo on creep properties of tows and
introduction to 3M Technologies and with 3M Technologies personnel on Nextel
fibers and the costs, and the library of University of Notre Dame for
the access of reference books.


\begin{figure}
 \plotone{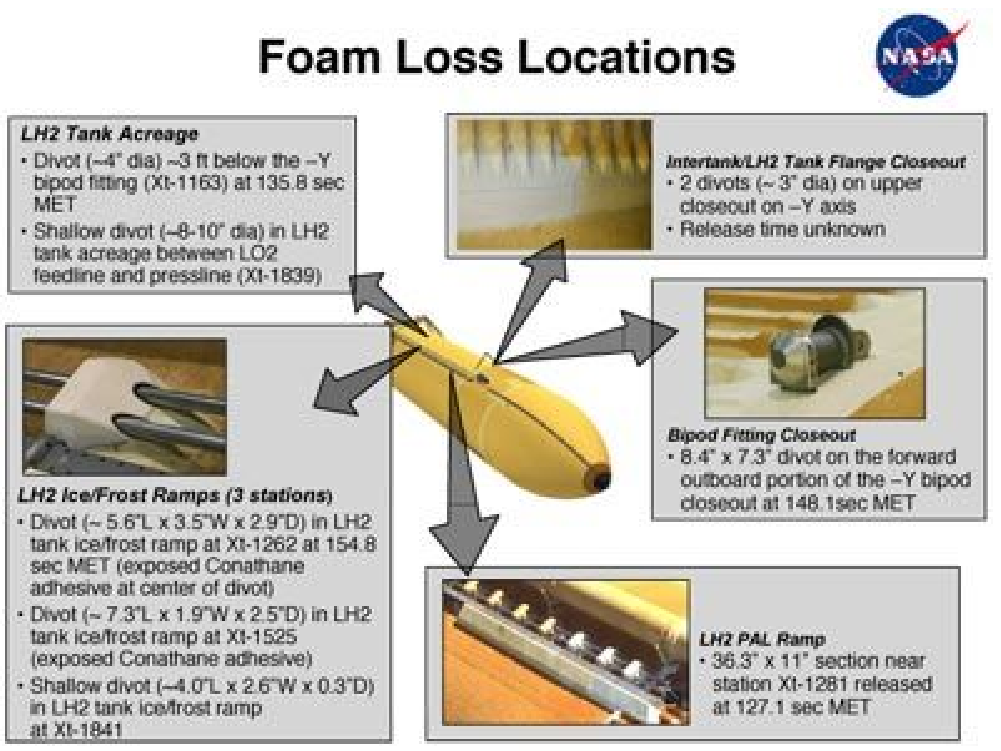}
 \caption{ Foam loss locations on STS-114 external fuel tank publicly
released by NASA: http://www.nasa.gov or http://www.spaceflightnow.com.}
\label{foamloss}
\end{figure}


\end{document}